\documentclass[10pt,amsmath,amssymb,latexsym]{article}
\usepackage{graphicx}  % needed for figures
\usepackage{dcolumn}   % needed for some tables
\usepackage{bm}        % for math
\usepackage{verbatim}  % for math

\newtheorem{myDefinition}{Definition}
\newtheorem{myPostulate}{Postulate}
\newtheorem{myProperty}{Property}

\newtheorem{myProof}{Proof}

\newcommand{\state}[1]{\left|{#1}\right>}
\newcommand{\stateBasis}[2]{\left|{#1}\right>_{#2}}
\newcommand{\bra}[1]{\left<{#1}\right|}
\newcommand{\braBasis}[2]{\left<{#1}\right|_{#2}}
\newcommand{\braket}[2]{\left<{#1}|{#2}\right>}

\begin{document}

\title{No quantum process can explain the existence of the preferred basis:\\decoherence is not universal}
\author{Hitoshi Inamori\\
\small\it Soci\'et\'e G\'en\'erale\\
\small\it Boulevard Franck Kupka, 92800 Puteaux, France\\
\small\it Previous academic affiliation: \\
\small\it Centre for Quantum Computation, Clarendon Laboratory, Oxford University
}

\bigskip

\date{\today}

\maketitle

\begin{abstract}
Environment induced decoherence, and other quantum processes, have been proposed in the literature to explain the apparent spontaneous selection -- out of the many mathematically eligible bases -- of a privileged measurement basis that corresponds to what we actually observe.

This paper describes such processes, and demonstrates that -- contrary to common belief -- no such process can actually lead to a preferred basis in general. The key observation is that environment induced decoherence implicitly assumes a prior independence of the observed system, the observer and the environment. However, such independence cannot be guaranteed, and we show that environment induced decoherence does not work in general. 

We conclude that the existence of the preferred basis must be postulated in quantum mechanics, and that changing the basis for a measurement is, and must be, described as an actual physical process.
 
\bigskip

\textbf{Keywords:} Quantum mechanics, Measurement, Preferred basis, Entanglement

\end{abstract}

\section{Introduction}  
Assume that an observer $O$ is performing a measurement on a quantum system $S$. For simplicity, suppose that $O$ and $S$ can both be described by Hilbert spaces of dimension two, and let $\{\stateBasis{i}{O}\}_{i=0,1}$ and $\{\stateBasis{i}{S}\}_{i=0,1}$ be some orthonormal bases for these spaces. Assume that the respective initial states of $O$ and $S$ are $\stateBasis{0}{O}$ and $\frac{1}{\sqrt{2}}\stateBasis{0}{S} + \frac{1}{\sqrt{2}}\stateBasis{1}{S}$.

The measurement is initiated by an interaction between the observer and the quantum system. Following~\cite{Zurek81, Zurek91, Zurek02}, suppose that the state of the system comprised of $O$ and $S$, denoted by $O\otimes S$, is transformed as below following this interaction:

\begin{equation}
\state{\phi_0} = \stateBasis{0}{O}\otimes\left(\frac{1}{\sqrt{2}}\stateBasis{0}{S} + \frac{1}{\sqrt{2}}\stateBasis{1}{S}\right)
\, \mapsto \,
\state{\phi_1} =\frac{1}{\sqrt{2}}\stateBasis{0}{O}\otimes\stateBasis{0}{S} + \frac{1}{\sqrt{2}}\stateBasis{1}{O}\otimes\stateBasis{1}{S}.
\end{equation}

One is tempted to interpret the state on the right hand side as describing a situation in which $O$ measures $S$ in the basis $\{\stateBasis{i}{S}\}_{i=0,1}\,$ : with probability $1/2$, $O$ sees $S$ in the state $\stateBasis{0}{S}$ and with probabilty $1/2$, $O$ sees $S$ in the state $\stateBasis{1}{S}$. 

The trouble is that the same resulting state can be written as well as~\cite{Zurek81, Zurek91,Zurek02}:

\begin{equation}
\state{\phi_1} =\frac{1}{\sqrt{2}}\stateBasis{+}{O}\otimes\stateBasis{+}{S} + \frac{1}{\sqrt{2}}\stateBasis{-}{O}\otimes\stateBasis{-}{S}
\end{equation}
where $\stateBasis{\pm}{O} = \frac{1}{\sqrt{2}}\left(\stateBasis{0}{O}\pm\stateBasis{1}{O}\right)$ and $\stateBasis{\pm}{S} = \frac{1}{\sqrt{2}}\left(\stateBasis{0}{S}\pm\stateBasis{1}{S}\right)$ constitute the conjugate bases for $O$ and $S$ respectively. Therefore the state $\state{\phi_1}$ could also be interpreted as a situation in which $O$ measures $S$ in the conjugate basis $\{\stateBasis{i}{S}\}_{i=+,-}$. So in which basis does $O$ observe $S$?

In practice, the actual observation is believed to be made in one basis and not in any other. If the experiment described is the Schr\"odinger cat experiment in which $O$ is the experimenter and $S$ is the cat, the cat is actually observed in the basis $\{\state{\textrm{live}}, \state{\textrm{dead}}\}$, never in the conjugate basis $\{\frac{1}{\sqrt{2}}\left(\state{\textrm{live}}+\state{\textrm{dead}}\right),\frac{1}{\sqrt{2}}\left( \state{\textrm{live}}-\state{\textrm{dead}}\right)\}$.

However the laws of quantum mechanics do not explicit tell us which basis is the preferred one in which actual observations are performed. This ambiguity is referred to as the \emph{preferred basis problem} in quantum mechanics.

The preferred basis problem has been at the centre of many studies and much debate~\cite{Schlosshauer05}.
In particular, theories such as \emph{environment induced decoherence}~\cite{Zeh70,Zurek81, Zurek91}, have been put forward in an effort to explain the spontaneous apparition of such preferred bases. The goal of these theories is to explain the emergence of these preferred bases using only the laws of quantum mechanics, usually through an interaction with a third auxiliary physical system~\cite{Zurek81, Zurek91, Wang14}  such as the environment. More precisely, these theories assert that the state of the observer and the measured system, like $\state{\phi_1}$ for $O\otimes S$ in the example above, evolves in a short frame of time into a \emph{classical mixture of states} as defined below:

\begin{myDefinition}
The combined system $O\otimes S$ is said to be in a \emph{classical mixture of states} if and only if there exist an orthonormal basis $\{\stateBasis{i}{O}\}_{i=1,2,\ldots}$ for $O$ and an orthonomal basis $\{\stateBasis{j}{S}\}_{j=1,2,\ldots}$ for $S$, such that the density matrix describing $O\otimes S$ can be written as:
\begin{equation}
\rho = \sum_{i,j} p_{i,j}\stateBasis{i}{O}\braBasis{i}{O}\otimes \stateBasis{j}{S}\braBasis{j}{S}
\end{equation}
for some nonnegative set of numbers $0\leq p_{i,j}\leq 1$, $i,j=1,2,\ldots$ such that $\sum_{i,j} p_{i,j}=1$.
\end{myDefinition}

A classical mixture of states corresponds to a classical probabilistic sum of outcomes, in which the state for $O$ and the state for $S$ are jointly distributed over the separable orthonormal basis $\{\stateBasis{i}{O}\otimes\stateBasis{j}{S}\}$ following a classical probability distribution $p_{i,j}$. Property~\ref{Property1} in the Appendix shows an example of a family of density matrices that are not classical mixture of states.

The aim of this paper is to describe the theories such as environment induced decoherence, starting with a simple case of the environment induced decoherence proposed in~\cite{Zurek81, Zurek91, Zurek02}, then generalizing to any theories that is supposed to lead to a classical mixture of states using only processes allowed by the laws of quantum mechanics. 

We observe that all these theories make an implicit assumption on the initial state of the combined system, comprised of the observed system, the observer, and any third auxiliary system introduced by such theories. 
However, such assumption cannot be guaranteed to hold, as we have no prior knowledge of what the quantum state of any physical system is. And indeed it can be proved that these processes do not lead to the emergence of any preferred basis for many initial states that are possible for the combined system. This is at odds with the belief commonly shared so far, as for instance in~\cite{Zurek81} which claims that environment induced decoherence systematically leads to a state for the observer and the quantum system~\cite{Zurek81p1519}.

We conclude that the existence of the preferred basis in quantum mechanics cannot be explained by quantum mechanics itself. The existence of the preferred basis must be a postulate, added to the existing laws of quantum mechanics.

The consequence of such a postulate is that any selection of the measurement basis -- other than the preferred one -- must be considered as an explicit, actual physical process. As any actual physical processes, the selection of this basis cannot be independent of the rest of the universe, depending on the initial state of the universe, state which is not known. The choice of the measurement basis cannot be proven to be independent of the system being observed, and this fact is made explicit by postulating the existence of the preferred basis  in the laws of quantum mechanics.

%First, even if there was such a process, a stubborn observer can still observe in a different basis, this is not forbidden by quantum laws in strict sense.

\section{Case study: environment induced decoherence does not lead to a classical mixture of states in general}

The environment induced decoherence~\cite{Zurek91,Zurek02} proposes to explain the apparition of a preferred basis as a consequence of an unavoidable interaction of the observer $O$ with a third physical system, called environment, denoted by $E$. 

It is argued that such interaction reduces in a short frame of time any quantum state describing $O\otimes S$ into a classical mixture of states: the latter describes a situation in which $O$ observes $S$ in a state chosen from an unique basis, with a certain classical probability distribution. This unique basis is deduced from the nature of the interaction with the environment, and corresponds to the preferred basis.

Coming back to our first example, following~\cite{Zurek91,Zurek02}, introduce the environment and assume that the initial state for the combined system $O\otimes S\otimes E$ is
\begin{equation}
\state{\psi_0} =\stateBasis{0}{O}\otimes\left(\frac{1}{\sqrt{2}}\stateBasis{0}{S} + \frac{1}{\sqrt{2}}\otimes\stateBasis{1}{S}\right)\otimes\stateBasis{0}{E},
\end{equation} 
where $\stateBasis{0}{E}$ is some state in the Hilbert space describing $E$.

As in the previous example, there is a first interaction between $O$ and $S$ leading to the state
\begin{equation}
\state{\psi_1} = \left(\frac{1}{\sqrt{2}}\stateBasis{0}{O}\otimes\stateBasis{0}{S} + \frac{1}{\sqrt{2}}\stateBasis{1}{O}\otimes\stateBasis{1}{S}\right)\otimes\stateBasis{0}{E}.
\end{equation}  

At this point, the environment induced decoherence states that there is an unavoidable interaction between $O$ and its environment $E$. Suppose that such an interaction can be written as 
\begin{equation}
\forall i,\quad\stateBasis{i}{O}\otimes\stateBasis{0}{E}\,\mapsto\, \stateBasis{i}{O}\otimes\stateBasis{i}{E}
\end{equation}  
\begin{eqnarray}
\forall i, & & \quad\stateBasis{i}{O}\otimes\stateBasis{0}{E}\,\mapsto\, \stateBasis{i}{O}\otimes\stateBasis{i}{E}\\
& & \quad\stateBasis{i}{O}\otimes\stateBasis{1}{E}\,\mapsto\, \stateBasis{i}{O}\otimes\stateBasis{\textrm{not}\, i}{E}
\end{eqnarray}
where $\{\stateBasis{i}{E}\}_{i=0,1}$ is an orthonomal basis which is associated with this interaction between $O$ and $E$. The state $\state{\psi_1}$ evolves into
\begin{equation}
\state{\psi_2} = \frac{1}{\sqrt{2}}\stateBasis{0}{O}\otimes\stateBasis{0}{S}\otimes\stateBasis{0}{E} + \frac{1}{\sqrt{2}}\stateBasis{1}{O}\otimes\stateBasis{1}{S}\otimes\stateBasis{1}{E}.
\end{equation}  

It is argued that any information in the environment is lost or ignored. The density matrix for the subsystem $O\otimes S$, $\rho_{OS}$, is obtained by tracing over the degree of freedom associated with $E$, namely:
\begin{equation}
\rho_{OS} = Tr_E\left[\state{\psi_2}\bra{\psi_2}\right] =\frac{1}{2} \stateBasis{0}{O}\braBasis{0}{O}\otimes\stateBasis{0}{S}\braBasis{0}{S}+\frac{1}{2} \stateBasis{1}{O}\braBasis{1}{O}\otimes\stateBasis{1}{S}\braBasis{1}{S}.
\end{equation}

Mathematically, the density matrix $\rho_{OS}$ above is a classical mixture of states and can be interpreted as describing a classical situation in which $O$ observes $\stateBasis{0}{S}$ with probability $1/2$ and $\stateBasis{1}{S}$ with probability $1/2$. 
  
The proponents of the decoherence theory assert that this is indeed what actually happens physically -- the form of the interaction with the environment has selected a preferred basis, in which the density matrix of the composite system $O\otimes S$ adopts a format that corresponds to a classical mixture of events, and that $S$ is actually observed in that preferred basis.
The physical setup describing the process is described in Figure~\ref{Fig1}.

\setlength{\unitlength}{0.025 in}
\begin{figure}[!h]
\centering
\begin{picture}(100,70)

\put(35,55){\circle*{4}}

\put(55,35){\circle*{4}}

\put(35,55){\line(0,-1){23}}

\put(35,35){\circle{6}}

\put(55,35){\line(0,-1){23}}

\put(55,15){\circle{6}}

\put(15,55){\line(1,0){60}}

\put(15,35){\line(1,0){60}}

\put(15,15){\line(1,0){60}}

\put(10,55){\makebox(0,0){$S$}}

\put(10,35){\makebox(0,0){$O$}}

\put(10,15){\makebox(0,0){$E$}}

\put(95,55){\makebox(0,0){\small measurement}}

\put(95,35){\makebox(0,0){\small measurement}}

\put(95,15){\makebox(0,0){\small ignored}}

\end{picture}
\caption{Setup describing the environment induced decoherence}\label{Fig1}
\end{figure}
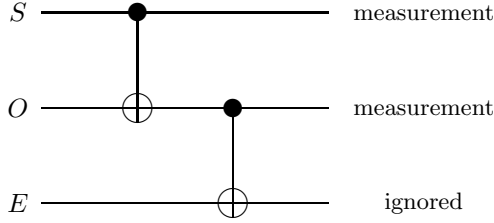

Our first remark is that even if the state of $O\otimes S$ is in a classical mixture of state in a given basis, nothing in the laws of quantum mechanics formally obliges the measurement to be done in such a basis. Rigourously speaking, quantum mechanics does not forbid observing the state $\rho_{OS}$ in a basis different from the basis in which $\rho_{OS}$ is diagonal.

This remark being set aside (although in the author's opinion, this is sufficient to call for the need to postulate the existence of the preferred basis), another fundamental issue with the decoherence theory is that the mechanism above ignores the possibility that the environment could have been entangled with the system or the observer previous to the experiment. The initial state of the whole setup $O\otimes S \otimes E$ is unknown, and in particular, the degree of the initial entanglement between the three systems is not known, and cannot be assumed.

As an example, suppose that the observed system and the environment were entangled, such that the initial state of $O\otimes S \otimes E$ is
\begin{equation}
\state{\psi'_0} = \frac{1}{\sqrt{2}}\stateBasis{0}{O}\otimes\stateBasis{0}{S}\otimes\stateBasis{0}{E} + \frac{1}{\sqrt{2}}\stateBasis{0}{O}\otimes\stateBasis{1}{S}\otimes\stateBasis{1}{E},
\end{equation}
then applying the same interactions above will lead to the final state
\begin{eqnarray}
\state{\psi'_2} &=& \frac{1}{\sqrt{2}}\stateBasis{0}{O}\otimes\stateBasis{0}{S}\otimes\stateBasis{0}{E} + \frac{1}{\sqrt{2}}\stateBasis{1}{O}\otimes\stateBasis{1}{S}\otimes\stateBasis{0}{E}\nonumber\\
&=&\left(\frac{1}{\sqrt{2}}\stateBasis{0}{O}\otimes\stateBasis{0}{S} + \frac{1}{\sqrt{2}}\stateBasis{1}{O}\otimes\stateBasis{1}{S}\right)\otimes\stateBasis{0}{E},
\end{eqnarray}

and tracing over $E$ leads now to
\begin{eqnarray}
\rho'_{OS} &=& Tr_E\left[\state{\psi'_2}\bra{\psi'_2}\right]\\
&=&\state{\chi}\bra{\chi}
\end{eqnarray}
where $\state{\chi}=\left(\frac{1}{\sqrt{2}}\stateBasis{0}{O}\otimes\stateBasis{0}{S} + \frac{1}{\sqrt{2}}\stateBasis{1}{O}\otimes\stateBasis{1}{S}\right)$. Such density matrix is not a classical mixture of states, as proved in Property~\ref{Property1}.

This example shows that the environment induced decohrence does not in general lead to a classical mixture state in a preferred basis: the initial state of the system is unknown and we cannot rule out entanglement between the observed system, the observer and the environment, which can lead to a final state in which we do not obtain a classical mixture of states.

\section{General case: no quantum process can lead to a classical mixture of states in general}

We have shown in the previous section that the environment induced decoherence as described in~\cite{Zurek91, Zurek02} does not always lead to a classical mixture of states, if one takes into account that the initial state of an experimental setup is ultimately unknown.

Is any other process, under the constraints of quantum mechanical laws, capable of producing classical mixture of states for the observed system and the observer, whatever the initial state is for the overall setup comprising the observed system, the observer, and any auxiliary third physical system?

However complex such a process may be, it can be described as follows: the overall setup is comprised of the three components, the observed system, $S$, the observer, $O$ and the auxiliary system, $E$. The process makes these three components interact, possibly in a most complex manner. After this interaction, represented by an unitary operator $U$,  we interest ourselves with the state describing $O\otimes S$, the state of the auxiliary system $E$ being ignored (Figure~\ref{Fig2}).

\setlength{\unitlength}{0.025 in}
\begin{figure}[!h]
\centering
\begin{picture}(100,70)

\put(15,13){\line(1,0){20}}
\put(22,10){\makebox(5,20){$\vdots$}}
\put(15,25){\line(1,0){20}}

\put(15,33){\line(1,0){20}}
\put(22,30){\makebox(5,20){$\vdots$}}
\put(15,45){\line(1,0){20}}

\put(15,53){\line(1,0){20}}
\put(22,50){\makebox(5,20){$\vdots$}}
\put(15,65){\line(1,0){20}}

\put(35,10){\framebox(20,60){$U$}}

\put(55,13){\line(1,0){20}}
\put(62,10){\makebox(5,20){$\vdots$}}
\put(55,25){\line(1,0){20}}

\put(55,33){\line(1,0){20}}
\put(62,30){\makebox(5,20){$\vdots$}}
\put(55,45){\line(1,0){20}}

\put(55,53){\line(1,0){20}}
\put(62,50){\makebox(5,20){$\vdots$}}
\put(55,65){\line(1,0){20}}

\put(5,60){\makebox(0,0){$O$}}

\put(5,40){\makebox(0,0){$S$}}

\put(5,20){\makebox(0,0){$E$}}

\put(80,65){\line(1,0){10}}

\put(80,33){\line(1,0){10}}

\put(90,65){\line(0,-1){32}}

\put(90,49){\line(1,0){5}}

\put(105,48){\makebox(0,0){$\rho_{OS}$}}

\put(95,20){\makebox(0,0){\small ignored}}

\end{picture}
\caption{Setup describing a general quantum process supposed to lead to classical mixture of states for $O\otimes S$}\label{Fig2}
\end{figure}
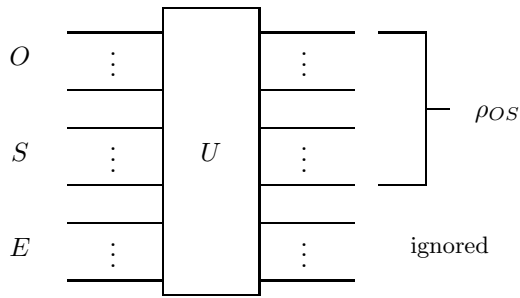

Let $\stateBasis{e}{E}$ be some state in the Hilbert space describing $E$. For any density matrix describing $O\otimes S$, $\rho_{OS}$, the initial state described by the density matrix:
\begin{equation}
\rho_0 = U^{\dagger} \left( \rho_{OS}\otimes \stateBasis{e}{E}\braBasis{e}{E} \right) U
\end{equation}
lead to the final state $U\rho_0 U^\dagger=\rho_{OS}\otimes \stateBasis{e}{E}\braBasis{e}{E}$. By tracing over $E$, we obtain naturally $\rho_{OS}$ for the final state of $O\otimes S$.

This in particular holds for any density matrix $\rho_{OS}$ that is not a classical mixture of states for $O\otimes S$. We have actually proven in Property~\ref{Property1} that many such density matrices exist. Therefore there exist initial states for the overall system $O\otimes S\otimes E$ that does not lead to a mixture state for $O\otimes S$.

We have therefore shown:

\begin{myProperty} 
In any measurement process involving an observer, an observed system and any auxiliary system, no physical setup which is obeying quantum mechanical laws can guarantee to produce a classical mixture of states for the observer and the observed system.
\end{myProperty}

\section{Consequences}
\subsection{The preferred basis must be postulated}
The result in the previous section shows that quantum mechanical processes cannot by themselves explain the existence of preferred bases in quantum mechanics. Current formulation of the quantum mechanical laws is not sufficient to imply the existence of a preferred basis. To be rigouroulsy complete, quantum mechanical laws must be supplemented by an explicit assumption on what the preferred basis is, at least for the space describing the observer.

\begin{myPostulate} The Hilbert space describing the observer is endowed with a preferred basis $\{\stateBasis{i}{O}^P\}_{i=1,2\ldots}$. Measurements are performed exclusively in this preferred basis. Given the reduced density matrix $\rho_O$ for $O$,  the probability that $O$ observes the result $i$ is given by $Tr\left[\rho_O\stateBasis{i}{O}^P\braBasis{i}{O}^P\right]$.
\label{postulate}\end{myPostulate}

Note that we need to assume the preferred basis for the observer only, as the existence of the preferred basis is experienced only at the observer's level. Changing the basis used for $S$ or $E$ in the calculation above has no impact on the outcomes and the associated probabilities experienced at the observer $O$'s level.

\subsection{Change of measurement basis is a physical process}
As quantum mechanics cannot explain the existence of preferred basis, we need to accept it as a postulate. There exists a privileged basis for the Hilbert space describing the observer, and as a consequence there is no theoretical freedom in the selection of the basis in which measurements are performed. Observation of quantum states can be done in different bases, but there must be an actual physical process corresponding to this basis change: 
for instance, in a photon polarisation measurement, the measurement basis is changed by actually acting on a phase shifter. In the Stern-Gerlach experiment the measurement basis is changed by rotating the magnets creating the magnetic field. Our point is that one does not change the measurement basis only by thought, and that an actual physical change must occur to act on a measurement basis.

The fact that a measurement basis change is not a theoretical concept but is an actual physical process has a fundamental consequence. So far, it has been common to assume (for instance, in the thought experiment leading to Bell inequality~\cite{Bell} in which two remote experimenters $A$ and $B$ choose freely and independently the measurement bases locally), that the choice of ``how'' a physical system is measured is independent of the state  of the measured physical system itself. This assumption seemed natural if one assumed a conceptual freedom in choosing the measurement basis.

By making the measurement basis selection an explicit physical process that also obeys to the laws of quantum mechanics, we have no longer the theoretical independence between an observed system and the measurement basis in which the system is observed. Indeed, the observed system and the setup selecting the measurement basis are both quantum systems and are described jointly by a composite quantum system, for which the initial state is unknown. As proved in~\cite{IsIndependentChoicePossible}, this implies that no mechanism can guarantee that the choice of the measurement basis is independent of the state of the system being measured. 

As such, Postulate~\ref{postulate} of the previous Section is not a mere mathematical axiom that is only required for a formal completeness of the laws of quantum mechanics. It implies a theoretical dependence between the choice of the measurement basis and the observed physical system, i.e. a potential \`a-priori dependence between the observer and the observed.

\appendix
\section{Appendix}
The following property gives an example of a family of density matrices that are not classical mixture of states, for any Hilbert spaces for $O$ and $S$:
\begin{myProperty}\label{Property1}
Let $\{\stateBasis{i}{O}\}_{i=1,2,\ldots}$ and $\{\stateBasis{j}{S}\}_{j=1,2,\ldots}$ be any orthonormal bases for the Hilbert spaces $O$ and $S$ respectively.
Consider any state $\state{\chi}\in O\otimes S$ such that
\begin{equation}
\state{\chi} = \sum_{i,j} \alpha_{i,j}\stateBasis{i}{O}\otimes \stateBasis{j}{S} \\
\end{equation}
with $\alpha_{i,j}$ being non-zero for at least for two couples $(i_1, j_1)$ and $(i_2, j_2)$, with $i_1\neq i_2$ and $j_1\neq j_2$. Then the density matrix $\rho = \state{\chi}\bra{\chi}$ is not a classical mixture of states.
\end{myProperty}

\begin{myProof}
The proof uses standard mathematical techniques as described for instance in~\cite{Mintert09}. 

Let's consider $Tr(\rho^2)$. On one hand, we have $\rho^2=\state{\chi}\braket{\chi}{\chi}\bra{\chi}=\state{\chi}\bra{\chi}$, therefore $Tr(\rho^2) = \braket{\chi}{\chi}=1$.

On the other hand, if we assume that $\rho$ is a classical mixture of states, then by definition, there exist an orthonormal basis $\left\{\stateBasis{\mu_i}{O}\otimes \stateBasis{\nu_j}{S}\right\}_{i,j}$ for $O\otimes S$, not necessarily equal to the basis $\left\{\stateBasis{i}{O}\otimes \stateBasis{j}{S}\right\}_{i,j}$, such that 
\begin{equation}
\rho = \sum_{i,j}p_{i,j}\stateBasis{\mu_i}{O}\braBasis{\mu_i}{O}\otimes\stateBasis{\nu_j}{S}\braBasis{\nu_j}{S},\label{App}
\end{equation}
with $0\leq p_{i,j}\leq 1$ and $\sum_{i,j}p_{i,j}=1$.

Taking the trace of its square, we get $Tr(\rho^2)= \sum_{i,j}p^2_{i,j}$ which can be equal to one only if there is an unique couple $(i_0, j_0)$ such that $p_{i_o,j_0}=1$, the remaining $p_{i,j}$ being zero.

Take now the trace over $S$. We have, on one hand, $Tr_S(\rho)=\sum_j \braBasis{j}{S}\big[\state{\chi}\bra{\chi}\big]\stateBasis{j}{S} = \sum_{i} \left(\sum_j |\alpha_{i,j}|^2\right)\stateBasis{i}{O}\braBasis{i}{O}$. This density matrix has a rank strictly greater than 1 as $\alpha_{i,j}$ is non-zero at least for two couples $(i_1, j_1)$ and $(i_2, j_2)$, with $i_1\neq i_2$ and $j_1\neq j_2$.

On the other hand, if we assume that $\rho$ is a classical mixture of states, then using Equation (\ref{App}), $Tr_S(\rho)=Tr_S\left(p_{i_o,j_0} \stateBasis{\mu_{i_0}}{O}\braBasis{\mu_{i_0}}{O}\otimes\stateBasis{\nu_{j_0}}{S}\braBasis{\nu_{j_0}}{S}\right)= \stateBasis{\mu_{i_0}}{O}\braBasis{\mu_{i_0}}{O}$ which is of rank 1. This is a contradiction, demonstrating that $\rho$ is not a classical mixture of states.
\end{myProof}
  
%\section{Acknowledgements}

\end{document}